\documentclass[pre,twocolumn,showpacs]{revtex4}

\usepackage{amsmath}
\usepackage{amssymb}

\usepackage{graphicx}

\renewcommand{\vec}[1]{{\bf #1}}
\newcommand{\evec}[1]{\vec{\hat #1}}
\newcommand{\mat}[1]{{\mathcal #1}}
\newcommand{\pot}[1]{{\mathcal #1}}

\begin{document}

\title {Hexagonal, square and stripe patterns of 
the ion channel density in biomembranes}

\author{Markus Hilt}
\author{Walter Zimmermann}
\affiliation{Theoretische Physik, Universit\"at Bayreuth, 
              D-95440 Bayreuth, Germany }

\date    {August 25, 2006}

\begin{abstract}
Transmembrane ion flow through channel proteins undergoing density fluctuations may cause lateral 
gradients of the electrical potential across the membrane giving rise to electrophoresis of charged 
channels. A model for the dynamics of the channel density and  the voltage drop across the 
membrane (cable equation)  coupled to 
a binding-release reaction with the  cell skeleton (P. Fromherz 
and W. Zimmerman, Phys.~Rev.~E {\bf 51}, R1659 (1995)) is analyzed  in one and two spatial dimensions. 
Due to the binding release reaction 
spatially periodic modulations of the channel density with a finite wave number 
are favored at the onset of pattern formation, whereby 
the wave number  decreases with the kinetic rate of the binding-release reaction.
In a two-dimensional extended membrane hexagonal modulations of the ion channel density are preferred 
in a large range of parameters. The stability diagrams of the periodic patterns near threshold 
are calculated and in addition 
the equations of motion in the limit of a slow binding-release kinetics are derived.
\end{abstract}


\maketitle

\section{Introduction}

Spatio-temporal pattern formation is ubiquitous in systems driven away 
from thermal equilibrium \cite{CrossHo,Cladis:95,Manneville:90,Murray:89}.
Many physical, chemical and biological systems display dissipative structures,
even though the underlying pattern forming mechanisms  are often completely different.
Nevertheless many of these patterns, especially those emerging at the primary 
bifurcation, belong to a few universality classes \cite{CrossHo} and
patterns occurring in rather disparate systems share qualitative and
unifying  properties.

Pattern forming processes in biological systems such as the fluid mosaic model, 
dilute filament-motor solutions (see e.g.
\cite{Nedelec:1997.1,Kardar:2001.1,Liverpool:2003.2,Ziebert:2005.2}), 
actively polymerizing filaments \cite{Ziebert:2004.1},
spiral waves in the cardiac system \cite{Panfilov:1995.1},
skin patterning of the angle fish \cite{Kondo:95.1}
or oscillatory dynamics in cell division 
\cite{Mandelkow:89.1,Mandelkow:90.3,Hammele:2003.1,Kruse:2005.1}, are in 
general more elaborate than in classical pattern forming systems as for instance in fluid dynamics \cite{CrossHo}. 
In the latter case the equations of motion can be derived by using elementary conservation laws 
and phenomenological transport laws and accordingly, for various patterns
in fluid dynamical systems a quantitative understanding
has been  achieved with a high precision \cite{Cladis:95,CrossHo}.
These achievements can serve as a guide for the analysis of more involved  biological
or chemical pattern forming systems, where the respective models cover only
the key steps of the complex biochemical reaction cycles.

In the present work we investigate pattern formation of ion channels embedded in a biomembrane.
Membranes are an important building block of living
cells and play a key role for the biological
architecture.  
They consist of a lipid bilayer which is rather impermeable and 
build the barrier to the cell environment. 
All the vital components needed inside the cell 
are transported across membranes through specific proteins.
Especially for the signal distribution along nerve cells (axons) 
the transport of ions through ion channels 
embedded in the membrane is essential since the 
transmembrane ion conductance is governed substantially by these
discrete channels. 
The channel proteins are considered to move freely along 
the fluid lipid bilayer which is referred to as 
the fluid mosaic model \cite{Singer:72.1}, a concept
that has attracted considerable attention. 
Describing the dynamics of ion channels within this framework,
one finds transitions to various stationary as well 
as time-dependent density patterns with possible biological implications 
\cite{Fromherz:88.1,Fromherz:88.2,Fromherz:91.1,Fromherz:94.1,Zimmermann:95.2,SKramer:2002.1,Leonetti:2005.1,RPeter:2006.100}. 
The binding-release reaction removes the conservation of
mobile ion channels and as a consequence causes pattern forming instabilities
with a finite wave number. Accordingly one expects 
in two-dimensional extended systems beyond a stationary bifurcation either stripes, squares or
hexagonal patterns as prototype patterns.  
Here we focus on the competition between these patterns 
in the presented model system.

Besides the fluid-mosaic model the channel concept \cite{Neher:76.1}
is the second central physical idea in the field of biomembranes. 
It was found that including their electro-diffusion properties 
\cite{Jaffe:77.1,Stollberg:88.1, Nitkin:90.1} ion channels 
have an intrinsic propensity for self-organization \cite{Fromherz:88.1}.
When a concentration gradient of 
salt across the membrane exceeds a certain threshold,
the conserved number of freely movable ion channels 
may organize into transient 
periodic patterns which finally decay into global clusters
\cite{Fromherz:88.2,Fromherz:91.1,Fromherz:94.1}.

The model of a fluid-mosaic of ion channels is
only elementary and  
neglects at least three important properties
of real biomembranes: (A) An interaction with signal 
molecules may induce a reversible
molecular transition which opens or closes an 
ion channel \cite{Neher:76.1}, (B) an
interaction with the cell skeleton may immobilize 
 ion channels \cite{Poo:86.1} and (C) the excluded volume interaction
between the ion channel molecules. 
The spatially dependent mobility of ion channels due
to rafts \cite{Simons:1997a} is also an effect
neglected in the fluid mosaic model, but this
heterogeneity effect is beyond the scope of the
present work.

The opening-closing reaction keeps the number of mobile ion channels 
conserved and its effects on the instability of the homogeneous ion channel distribution
have been investigated thoroughly in two recent publications
\cite{SKramer:2002.1,RPeter:2006.100}. 
Since membrane deformations coupled to the underlying cell skeleton ({\it actin cortex}) 
may also open or close ion channels \cite{Aidley:96,Guharay:84.1}, additionally
we take here 
both the immobilization and the closing of the channels into account \cite{Zimmermann:95.2}.
For the sake of simplicity we choose a model which combines
the two processes: We consider a reversible binding-release 
reaction of ion channels with the cell skeleton
and assume that this interaction induces a closing of the channels.
Alternative models for pattern formation along the
cell membrane take additional 
intermediate steps of the opening-closing dynamics into account
\cite{Fromherz:88.2,SKramer:2002.1,Korogod:2000.1,Leonetti:2005.1}
but a thorough analysis of the pattern formation processes 
in two spatial dimensions is not available yet.

In the model we propose the closed ion channels which are considered to be bound
to the cell skeleton are acting as a source for the mobile and open channels
and therefore, the free and open ion channels are not conserved anymore in contrast to
previously discussed models.
We find in this model different kinds of selforganization 
of the mobile ion channels:  With a considerable
binding-release reaction one has an unconserved number 
of open channels and one finds (a) stable stripe or
hexagonal patterns  above threshold. This formation of
stationary periodic
patterns belongs to the same universality class as, 
for example, convection rolls in
hydrodynamic systems. 
(b) The transition into the periodic pattern is either sub- or supercritical
depending on the equilibrium constant, the relaxation 
time of the binding-release
reaction and also on the strength of the excluded volume interaction of the ion channel molecules.
In the limit of small binding-release reaction rates 
the model shows a crossover between pattern formation
for an unconserved and a conserved order parameter.
For this crossover regime a reduced equation is derived in 
Sec.~\ref{weakspin}. 

This work is organized as follows:
In Sec.~\ref{modelsys} we describe the model system and we give the basic equations for the
analysis in the  subsequent sections.
The linear stability of the homogeneous
distribution of the ion channel density 
and the onset of the patterns is discussed in Sec.~\ref{onset}.
The amplitude equations, describing 
the weakly nonlinear behavior of stripes, hexagonal 
and square patterns are derived in  Sec.~\ref{weaknon}.
In this section also the nonlinear competition between
these patterns is investigated by a thorough analysis.
In Sec.~\ref{numresults} numerical solutions of the
model equations are presented. Those numerical results
provide an estimate for the validity range
of the perturbational analysis given in 
Sec.~\ref{weaknon}. Concluding remarks and a
discussion of the results are provided by Sec.~\ref{conclusion}.

\section{Model System}
\label{modelsys}

We consider a model membrane with embedded ion channels separating a thin electrolytic layer
from an electrolytic bulk medium. This may refer to a cell membrane in close contact to another
cell or to a membrane cable as it occurs in dendrites and axons of neurons. In the first case
the thin layer is given by the extracellular cleft and the bulk by the cytoplasm. A particularly
important biological example of this case is the post synaptic membrane of a neuronal synapse.
In the second case the narrow cylindrical cytoplasm plays the role of a one-dimensional cleft
opposed by the extracellular bulk medium.

\begin {figure}[hbt]
\includegraphics[width=8cm]{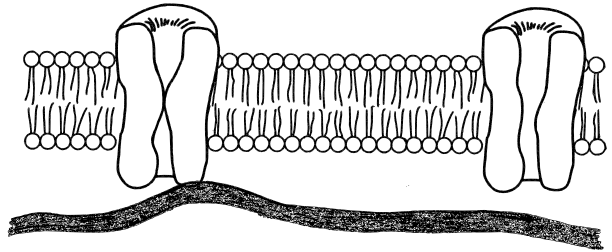}
\caption { 
A fluid membrane that separates a narrow cleft of electrolyte from a bulk 
electrolytic phase is considered. Membrane proteins are 
embedded in the lipid bilayer. They are mobile
along the membrane (diffusion coefficient $D$), they form selective
ion channels across the membrane (conductance $\Lambda$),
they bear an electrophoretic charge $q$ and they interact with
a filamentous substrate of the membrane (cell skeleton) via a
binding-release reaction with rate constants $\kappa_o$ and $\kappa_c$.
Binding closes the channels by a conformational change. The 
system is driven by a concentration gradient of those ions 
which are conducted by the channels (Nernst-type 
potential $E$). The  model refers to the
biological situations of a cell-cell-contact (post-synaptic
membrane of a synapse) and of a cylindric cellular 
cable (neuron dendrites). The cleft of the model corresponds to
the extracellular space in the first case,
whereas it corresponds to the narrow cytoplasm in the second case. 
The structure of the model is described by the density of mobile
channels $n(\vec{r},t)$ and by the voltage in the cleft $v(\vec{r},t)$ as a
function of space $\vec{r}$ and time $t$.
}
\label{fig1}
\end{figure}

Here we assume that the ion channels interact with the 
underlying cell skeleton, cf. Fig.~\ref{fig1}
and Refs.~\cite{Aidley:96, Guharay:84.1, Fromherz:91.2}.
In a reversible binding-release reaction 
among the ion channels and the cell skeleton the
ion channels undergo also a conformational change 
and switch between an opened and a closed state.
This binding-release reaction of the ion channels (IC) 
is described by a simple reaction scheme 
with two rate constants $\kappa_o$ and $\kappa_c$, 
i.e. by an equilibrium constant 
$K_{BR} = \kappa_o/\kappa_c$ and a  relaxation 
time $\tau_{BR} = (\kappa_o + \kappa_c)^{-1}$:
\begin{align}
\label{reacequi}
IC^{bound}_{closed}
\begin{array}{c} \kappa_o \\ \rightleftharpoons \\ \kappa_c \end{array}
IC^{free}_{open}
\end{align}
The closed ion channels are bound  to the cell skeleton and represent a source for free and open 
channels via the binding-release reaction. 
Accordingly the number of open and free channels is not conserved in our model, whereas in several 
previous investigations the number of free and open 
channels was conserved. Further mechanisms with similar consequences as the nonconservation of the
ion channel number are also known \cite{Korogod:94.1}.

The free ion channels with an electrical conductance $\Lambda$ undergo a Brownian motion along the 
membrane with a diffusion coefficient $D$. They are selective for ions presenting a concentration
gradient across the membrane which is described by a Nernst-type potential $E$.
The proteins bear an effective electrophoretic charge $q$ leading to a drift motion in a lateral 
electrical field.
The current through the mobile and open channels and through a homogeneous leak conductance of the 
membrane affects the local voltage in the cleft. In combination with an inhomogeneous distribution 
of the ion channels this gives rise to lateral gradients of the voltage.

\subsection{Basic equations}
In a mean field approximation the local density $n(\vec{r},t)$ of free and open channels (particles per 
unit area) is determined by  diffusion and electrophoretic drift of the ion channels, the binding-release
reaction and the local interaction forces.
The homogeneous density $\overline n$ is kept constant by a reservoir $\overline n_c =const.$ of 
bound channels by an equilibrium binding-release reaction given by Eq.~(\ref{reacequi}) with
$\overline n = K_{BR} \overline n_c$.
The equations of motion for the channel density may be expressed in terms of the deviation
\begin{align}
\tilde n = n - \overline n
\end{align}
from the mean density $\overline n$, where the dynamics is composed of a lateral current gradient and a
source-sink contribution
\begin{align}
\partial_t \tilde n \ = - \nabla \cdot \vec{j}(n) - \frac{\tilde n }{\tau_{BR}}\, .
\end{align}
The current depends on the  part of the chemical potential, which is  independent of the electric field, and
on the mobility of the  charged channels times the local electric field
\begin{align}
\vec{j}(n) = -\nabla \mu  -\nu q n  \nabla v 
\end{align}
with the mobility
\begin{align}
\nu = \frac{D}{k_B T}\enspace,
\end{align}
the diffusion constant $D$, the effective charge $q$ per channel, and the local voltage drop 
across the membrane $v(\vec{r},t)$.
The charge $q$ is assumed to be an effective charge taking into account screening as well as 
electro-osmotic effects \cite{Leonetti:1998.1}.
The remaining part of the chemical potential may be derived from the free energy
\begin{align}
\mu = \frac{\delta \pot{F}} {\delta \tilde n}\enspace,
\end{align}
where the free energy (per unit area S) of the channel-channel interaction is up to leading 
order in the deviation $\tilde n$ of the following from 
\begin{align}
\label{freeen}
\pot{F}= \frac{1}{S} \int_S d\vec{r}
\left[\,
\frac{D}{2}   \,{\tilde n}^2  + \,
\frac{g_2}{3} \,{\tilde n}^3  + \,
\frac{g}{4}   \, {\tilde n}^4 + \,
\frac{1}{2} \xi^2 (\nabla {\tilde n})^2 \,
\right]\,.
\end{align}
The higher order contributions become important especially for a negative diffusion constant $D$, 
if a demixing between lipid proteins and  ion channels in the membrane 
takes place. 
Since the diffusion constant is assumed to be always positive for the present problem, the higher 
order contributions may be neglected in most situations.
However, if the amplitudes of the spatial ion channel density modulations, as induced by the pattern 
forming mechanism discussed in this work, become strong, the second and third order terms in 
Eq.~\eqref{freeen}, describing the effects of excluded volume interactions become important in some 
range of parameters. 
Here we investigate exemplarily the stabilizing effects of the fourth order contribution to 
Eq.~\eqref{freeen}, i.e. with $g_2=0$, $\xi=0$ but $g \ne 0$. 
In this  case the equation of motion for $\tilde n$ takes the following form
\begin{align}
\label{201}
\partial_t \tilde n =&
\nabla^2 \left(D {\tilde n} +g \tilde n^3 \right)
+   \frac{q D}{k_B T} \, \nabla \cdot \left[ \left (\tilde n + \overline n\right) \nabla v \right]
-\, \frac{\tilde n }{\tau_{BR}}\, .
\end{align}
The voltage $v$ in the cleft is obtained from Kirchhoff's law for each element of the membrane 
cable.
Taking into account the current across the membrane and along the core of the cable or 
below a flat membrane, we obtain the Kelvin equation with the membrane capacitance $C$ either per unit 
length in a one dimensional model or per unit area for a two dimensional model, respectively, the resistance $R$ of the cleft per unit length (area) and the leak 
conductance $G$ of the membrane per unit length (area)
\cite{Scott:75.1,Noble:1975}
\begin{align}
\label{202}
C \partial_t v = \frac{1}{R} \nabla^2 v - G\,v - \Lambda n (v -E)\,.
\end{align}
The  special case $ \tau_{BR} \rightarrow \infty$ and $g=0$ of these equations has been investigated in
Refs.~\cite{Fromherz:88.1,Fromherz:91.1,Fromherz:94.1}.

\subsection{Scaled Equations}
\label{secscaleq}

We rescale Eq.~(\ref{201}) and Eq.~(\ref{202}) by introducing dimensionless coordinates for space
$\vec{r}' \,=\,\vec{r} / \lambda$ and time $t'\,=\,t\,/\,\tau$ with the typical length scale of
an electrical perturbation $\lambda \,=\, [R (\Lambda {\overline n} +G)]^{-1/2}$ and the time
constant of displacement $\tau = \lambda^2/D$.
We use normalized variables for the particle density $N = (n-\overline n) / \overline n $ and
voltage $V = (v-v_R)q/k_BT$ with the  resting voltage $v_R = \alpha E$ and the density parameter
$\alpha = \Lambda \overline n / (\Lambda \overline n +G )$.
Introducing the normalized relaxation time $\tau_V = R C D$ we obtain the normalized reactive
Smoluchowski--Kelvin equations \cite{Leonetti:1998.1}
\begin{subequations}
\label{scaleq}
\begin{align}
\label{scaleq1}
\partial_{t'} N
=& \nabla'^2 \left[ N +  g N^3 \right] + \nabla' \cdot \left[(1+ N) \nabla' V\right]-\beta N \, ,  \\
\label{scaleq2}
\tau_V \partial_{t'} V
=& \left[ \nabla'^2 -1 \right] V - \alpha (1-\alpha)\, \varepsilon  N - \alpha N V\enspace.
\end{align}
\end{subequations}
The  dynamics of the system  is controlled by the following three parameters:
(i)   The density parameter $\alpha$ characterizes the equilibrium of the binding-release
      reaction;
(ii)  the rate parameter $\beta = \tau / \tau_{BR}$ which characterizes the dynamics of the 
      binding-release and the simultaneous opening-closing reaction;
(iii) the control parameter $\varepsilon= - q E/( k_B T)$ which characterizes the distance to thermal equilibrium.
Since the spread of the voltage is fast compared to the diffusion of ion channels 
($ R \approx 10^8 \Omega, C \approx 1 \mu F/cm^2, D\approx 0.1 \mu m^2/s \rightarrow \tau_v \ll 1$ 
\cite{Hille:92})
we put $\tau_V=0$ in the following.
For simplicity the primes of the new coordinates $\vec{r}'$ and $t'$ are suppressed further on.

\section{The onset of periodic patterns}
\label{onset}
\begin{figure}[hbt]
\begin{center}
\includegraphics[width=0.95\columnwidth]{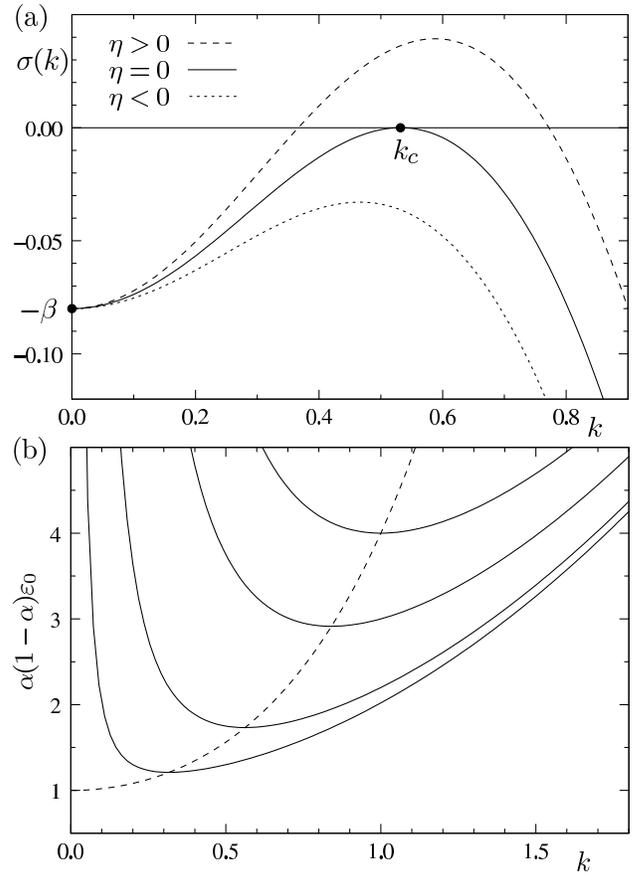}
\end{center}
\caption{
Part (a) shows the dispersion relation $\sigma(k)$ as given by Eq.~(\ref{302}) for a 
supercritical, a critical and subcritical value of the reduced control parameter $\eta$ 
and for the rate parameter $\beta=0.08$.
Part (b) shows for four different values of the rate parameter $\beta= 0.01,0.1,0.5,1.0$ 
(bottom to top) the neutral curves $\alpha(1-\alpha)\varepsilon_0(k)$ (solid lines) 
with $\varepsilon_0(k)$ given by Eq.~(\ref{305}). The dashed curve in (b) marks the 
location of the minimum of these neutral curves, i.e. the critical wave number 
$(k_c,\varepsilon_c)\, (\beta)$ as given by Eq.~(\ref{306a}) and Eq.~(\ref{306b}).
}
\label{figthresh}
\end{figure}

The onset of spatial patterns takes place in a parameter range, where the homogeneous density
$n=\overline n$ of the mobile channels, i.e.  $N=0$ and $V=0$, becomes linearly unstable with 
respect to small inhomogeneous perturbations.
In order to calculate this instability  the linear part of Eqs.~(\ref{scaleq}) is transformed 
by an ansatz
\begin{align}
\left( \begin{array}{c} N \\ V \end{array} \right) =
\left( \begin{array}{c} \delta\! N \\ \delta\! V \end{array}\right)
 e^{\sigma t + i \vec{k}\vec{r}}
\end{align}
into linear algebraic equations. The solubility condition of these equations determines 
for finite perturbations $\delta N, ~\delta V \not =0$ and for $\tau_V=0$ the dispersion 
relation $\sigma(k)$ 
\begin{align}
\label{302}
\sigma  = -k^2 \left( 1 - \frac{ \alpha (1-\alpha) \varepsilon}{1 + k^2 } \right)
-\beta \,\,
\end{align}
with $k=|\vec{k}|$, which is shown for three different values of the 
control parameter $\varepsilon$ in Fig.~\ref{figthresh}(a). A perturbation grows in the range
of the wave number $k$ where the real quantity $\sigma(k)$ becomes positive.  
The {\it neutral stability condition} $\sigma(k)$=0 applied to the expression in Eq.~(\ref{302}) 
gives the neutral curve as follows
\begin{align}
\label{305}
\varepsilon_0(k)  =  \frac{1+k^2}{\alpha (1-\alpha)} \, \left(1+ \frac{\beta}{k^2} \right)\,  ,
\end{align}
where $\varepsilon_0(k)$ separates the range of stable from the unstable parameter values.
A set of neutral curves $\varepsilon_0(k;\alpha,\beta)$ is shown in Fig.~\ref{figthresh}(b) 
for different values of the rate parameter $\beta$.

The minimum of $\varepsilon_0(k;\alpha,\beta)$ defines the critical wave number 
\begin{align}
\label{306a}
k_c = \beta^{\frac{1}{4}}
\end{align}
and the critical control parameter
\begin{align}
\label{306b}
\varepsilon_c = \frac{\left( 1+\sqrt{\beta} \right)^2}{\alpha (1-\alpha)}
\end{align}
at which the basic state becomes first unstable.

For $\varepsilon$-values above the neutral curve $\varepsilon_0(k,\alpha, \beta)$ the  growth 
rate $\sigma$ 
is positive and it takes its maximum at the wave number $k_m$
\begin{align}
k_m^2 = -1 + \left( 1+\sqrt{\beta} \,\right) \, \sqrt{1+\eta} \,,
\end{align}
whereby the reduced control parameter $\eta$
\begin{align}
\label{eta}
 \eta = \frac{\varepsilon-\varepsilon_c}{\varepsilon_c} \,\,
\end{align}
has been introduced. 
For a vanishing rate parameter $\beta=0$ the number of channels is conserved and  the dispersion in 
 Eq.~(\ref{302}) is at small values of $k$ proportional to $k^2$ which is similar to 
hydrodynamic excitation modes.
To some extent this limit has already been investigated previously
\cite{Fromherz:88.1,Fromherz:88.2,Fromherz:91.1}.

\section{Weakly nonlinear analysis}
\label{weaknon}

For a finite value of the binding-release reaction parameter $\beta $ the fraction of the
open ion channels is not conserved. 
In addition the homogeneous distribution of the ion channel density as well as the homogeneous 
voltage drop across the membrane may become simultaneously unstable against perturbations above 
a certain threshold.
The perturbation with the wave number $k_c=\beta^{1/4}$ has the largest growth rate, 
as described in the previous section.

Near threshold and in two spatial dimensions 
this periodic instability may lead to
stripe, square and, if the up-down symmetry is broken, 
also to  hexagonal patterns \cite{CrossHo}.
In a parameter range where the amplitudes 
of these patterns  are still small, 
the slow spatial variations of these patterns 
 may be described  
in terms of generic amplitude equations, 
a method used for many other physical, chemical and biological 
pattern forming systems \cite{CrossHo,Newell:92.1}.
These generic equations and their corresponding functionals are derived in this section from 
the present model system whereby the detailed scheme of derivation 
is given exemplarily for stripes in App.~\ref{appA}.
The parameter ranges where each pattern realizes the lowest functional value and
where two 
patterns coexist are determined in Sec.~\ref{compet}. 
Far beyond the threshold the solutions are determined numerically and the question of 
preference of patterns is addressed by numerical simulations in
Sec.~\ref{numresults}.

In the range where the binding-release
reaction becomes rather slow and the number of channels nearly conserved, 
i.e.  $\beta \propto \eta^2$, another set of equations is
 derived and discussed in part~\ref{weakspin}.
The analytical solutions derived in the two cases 
 $\beta \propto O(1)$ and   $\beta \propto \eta^2$  
are compared with numerical solutions of Eqs.~\eqref{scaleq}
in Sec.~\ref{numresults}.

\subsection{Periodic patterns for a finite binding-release reaction, $\beta \propto O(1)$}
\label{weakper}

\subsubsection{Stripe patterns}
\label{onedipat}
For finite values of $\beta$ the solution of the linear part of Eqs.~(\ref{scaleq}) is in
the simplest case spatially periodic in one direction. The two fields $N$ and $V$
may be written as a vector $\vec{u}=\left(N,V \right)$ in the form
\begin{align}
\label{fieldamp1}
\vec{u}_0 =
\evec{e}_0\,A \,e^{i \vec{k}_c \vec{r}}  + c.c.
\end{align}
with the eigenvector 
\begin{align}
\evec{e}_0 = \left( \begin{array}{c}
1 \\ E_0
\end{array}
\right), \quad E_0 = -\left(1+\sqrt{\beta} \right)
\end{align}
where c.c. denotes the complex conjugate of the preceding term.
Due to rotational symmetry the orientation of the wave vector
\begin{align}
\vec{k}_c = k_c \left(\begin{array}{c}
1 \\ 0
\end{array} \right)
\end{align}
may be chosen parallel to the x-direction. Spatial variations of the pattern which are 
slow on the length scale $2 \pi/k_c$, are described in Eq.~\eqref{fieldamp1} by a spatially 
dependent amplitude $A(x,y,t)$.
The evolution equation for this spatially and time dependent amplitude is a so-called 
amplitude equation, namely the Ginzburg-Landau equation in our case,
which takes for an isotropic two-dimensional system, as for instance for the 
two-dimensional flat membrane,
the following universal form \cite{Newell:92.1,Manneville:90,CrossHo}:
\begin{align}
\label{amplitudeeq}
\tau_0 \partial_t A  =
\left[ \eta   + \xi_0^2 \left(\partial_x - \frac{i}{2 k_c}\partial_y^2 \right)^2
- \gamma |A|^2 \right] A \, .
\end{align}
The values of the coefficients $\tau_0$, $\xi_0$ and $\gamma$ depend on the specific pattern 
forming system \cite{Manneville:90,CrossHo,Zimmermann:91.1}.
The relaxation time $\tau_0$ and the coherence 
length $\xi_0$ in the amplitude equation (\ref{amplitudeeq})
may be derived  from the dispersion relation $\sigma(k)$
in Eq.~(\ref{302}) and from the neutral curve $\varepsilon_0(k)$ 
specified in Eq.~\eqref{305} by a Taylor
expansion of both formulas around the critical point $(k_c,\varepsilon_c)$
(see for instance Refs.~\cite{Brand:86.1,CrossHo,Zimmermann:93.1}):
\begin{align}
\label{dispneu}
\tau_0^{-1} &= \left.\frac{\partial \sigma}{\partial \varepsilon}\right|_{\varepsilon=\varepsilon_c} \, \varepsilon_c~,
&
\xi_0^2 &= \frac{1}{2 \varepsilon_c} \frac{\partial^2 \varepsilon_0}{\partial k^2} \,.
\end{align}

Using the expressions in Eq.~(\ref{302}) and Eq.~(\ref{305}) we obtain their explicit form for 
the present model:
\begin{align}
\label{431}
\tau_0 &=\frac{1}{\sqrt{\beta} \left(1+ \sqrt{\beta} \right)}\,,
&
\xi_0^2 &= \frac{4}{\left(1+ \sqrt{\beta} \right)^2}\,.
\end{align}
The sign of the nonlinear coefficient $\gamma$ determines, whether the
transition  to the  periodic state as described by Eq.~(\ref{fieldamp1})  
is supercritical ($\gamma >0$) 
or subcritical ($\gamma <0$),
and it may be derived  by a perturbation calculation from the basic equations (\ref{scaleq}): 
\begin{align}
\label{ngamma}
\gamma =& \frac{3 g}{1+\sqrt{\beta}}
- \frac{1}{3} \Biggl[\, \frac{6\alpha^2-\left(2+2\sqrt{\beta}-\alpha\right)^2}{1+\sqrt{\beta} }
 \\
&\quad+\frac{2}{3\sqrt{\beta}} \left(4\sqrt{\beta}-2\alpha+1\right)\left(\sqrt{\beta}-2\alpha+1\right)
 \, \Biggr]\,.\nonumber 
\end{align}
The common scheme for the derivation of $\gamma$ may be found for instance in 
Refs.~\cite{CrossHo,Newell:92.1,Zimmermann:93.1} and the details of this calculation for
the present system are given in App.~\ref{appA}.

The linear coefficients in Eq.~\eqref{431} depend only on the rate parameter $\beta$.
By increasing the  binding-release reactions the relaxation time and also the correlation length 
becomes smaller.
The nonlinear coefficient depends besides the  rate parameter $\beta$ also  on the density parameter 
$\alpha$ and on the nonlinear interaction parameter $g$.
In the limit $\beta \to 0$ for $\alpha \neq 1/2$ the relaxation time $\tau_0$ and the nonlinear 
coefficient $\gamma$ diverge.
This behavior reflects the fact that 
in this limit the validity range of the amplitude equation is left and a different
perturbation expansion has to be used as described in Sec.~\ref{weakspin} below.

The amplitude equation (\ref{amplitudeeq}) for a stripe solution can also be derived from a 
functional $\pot{F}$
\begin{align}
\label{amplistripe}
\tau_0 \partial_t A = - \frac{\delta \pot{F}_S}{\delta A^{\ast}} \,,
\end{align}
($A^\ast$ denotes the complex conjugate of $A$) of the following form
\begin{align}
\label{potstripe}
\pot{F}_S =\frac{1}{S}
\int_S d\vec{r} \left[ \frac{\gamma}{2} |A|^4 -\eta |A|^2
+ \xi_0^2 \left|\left(\partial_x - \frac{i}{2k_c}\, \partial_y^2 \right)A\right|^2
\right].
\end{align}
In the following we will focus on spatially homogeneous patterns and their competition,
i.e. $A(x,t) \rightarrow A(t)$. So the Ginzburg-Landau equation \eqref{amplitudeeq}  
reduces to a Landau equation
\begin{align}
\label{amplitudeeq2}
\tau_0 \partial_t A  = \eta A - \gamma |A|^2  A 
\end{align}
with a simplified functional
\begin{align}
\label{potstripe2}
\pot{F}_S 
=\frac{1}{S}\int_S d\vec{r} \left[\frac{\gamma}{2} |A|^4 -\eta |A|^2\right]
=\frac{\gamma}{2} |A|^4 -\eta |A|^2\,.
\end{align}
Besides the trivial solution $A=0$, Eq.~\eqref{amplitudeeq2} has a second stationary solution 
\begin{align}
\label{stripesol}
A=\sqrt{\frac{\eta}{\gamma}}\,.
\end{align}
This solution exists in the supercritical case $\gamma>0$ only above the threshold and for $\gamma<0$
only in the range $\eta<0$ on the unstable branch of the subcritical bifurcation. As the herein presented
expansion breaks down for subcritically bifurcating stripes, higher order terms with respect to $A$ would
have to be taken into account in order to achieve a limitation of the amplitude $A$ which may however
be determined by solving Eqs.~(\ref{scaleq}) beyond threshold numerically as done in
Sec.~\ref{numresults}.

\begin {figure}[hbt]
\includegraphics[width=0.95\columnwidth]{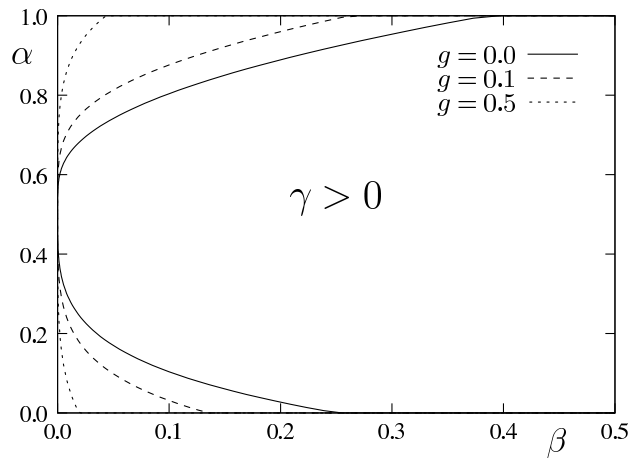}
\caption{
The lines describe the tricritical point, i.e.  $\gamma(\alpha,\beta,g)=0$,
for  the stripe pattern in the $\alpha-\beta$ plane and for different values 
of the interaction coefficient $g$, respectively. On the right hand side of each curve 
corresponding to different values of $g$, $\gamma$ is positive and stripes bifurcate supercritically.
}
\label{figbif}
\end{figure}

The parameter range  of the supercritical and  subcritical bifurcation are separated by the tricritical
line $\gamma(\alpha, \beta)=0$. This line is shown in Fig.~\ref{figbif} in the $\alpha-\beta$ plane 
for different values of the excluded volume parameter $g$, where the supercritical range may be 
extended by increasing the nonlinear parameter $g$.

In the limit $\beta \to 0$ and $\alpha \to 1/2$ the nonlinear coefficient $\gamma = 3 g + 1/4 > 0 $
is positive and the bifurcation still supercritical. Otherwise $\gamma$ diverges in the limit $\beta \to 0$.
This is also in agreement with an alternative perturbation analysis as described
in Sec.~\ref{weakspin}.

\subsubsection{Square patterns}
\label{squarepatt}
Square patterns can be described by a superposition of two
periodic waves as given by Eq.~(\ref{fieldamp1})
\begin{align}
\label{squareansatz}
\vec{u}_0
=\evec{e}_0\,&\left(A_1 e^{i \vec{k}_1 \vec{r}} + A_2 e^{i \vec{k}_2 \vec{r}} \right) + c.c. \, ,
\end{align}
whereby the two wave vectors $\vec{k}_1 $ and $\vec{k}_2 $ have the same length but both
are orthogonal to each other 
\begin{align}
\vec{k}_1   = k_c           \left(\begin{array}{c}  1 \\ 0 \end{array} \right)
\quad{\rm and}\quad
\vec{k}_{2} = k_c \left(\begin{array}{c} 0 \\ 1  \end{array}\right)\,.
\end{align} 
Similar as for stripes, one may derive by a perturbation calculation from the basic 
Eqs.~(\ref{scaleq}) the following two coupled equations for the two amplitudes $A_1$ 
and $A_2$
\begin{subequations}
\label{amplisquare}
\begin{align}
\tau_0 \partial_t A_1 =& \,\eta\, A_1  -
\left( \gamma\, |A_1|^2 + \chi\, |A_2|^2  \right)A_1\,, \\
\tau_0 \partial_t A_2 =& \,\eta\, A_2  -
\left( \gamma\, |A_2|^2 + \chi\, |A_1|^2  \right)A_2\,, 
\end{align}
\end{subequations}
wherein the spatial dependence of the amplitudes has already been discarded.
Herein $\tau_0$ and  $\gamma$ are defined by the same expressions as for stripes in 
Eq.~\eqref{431} and Eq.~\eqref{ngamma}. 
For squares one obtains with a perturbation calculation, similar as described in App.~\ref{appA}, 
the same expression for $\gamma$ as in Eq.~\eqref{ngamma} and the nonlinear coupling term $\chi$
\begin{align}
\label{nchi}
\chi =&
-4\left( 2(1-2\alpha)+\sqrt{\beta} \right)\\
&+\frac{2\left(3g+\alpha^2-2(1-2\alpha)^2 \right)}{1+\sqrt{\beta}}
-\frac{4(1-2\alpha)^2}{\sqrt{\beta}(1+\sqrt{\beta})} \enspace.
\end{align}
As for stripes, the two coupled equations may be once again be derived from a functional
\begin{align}
\tau_0 \partial_t A_i = - \frac{\delta \pot{F}_Q}{\delta A_i^* }
\end{align}
by determining the extremal value of the functional
\begin{align}
\label{potsquare}
\pot{F}_Q =
\sum_{i=1}^2 \left( \frac{\gamma}{2}\,|A_i|^4 - \eta\,|A_i|^2 \right)
+ \chi  |A_1|^2\, |A_2|^2\,.
\end{align}
Apart from the trivial solution $A_1=A_2=0$,
the coupled amplitude equations $(\ref{amplisquare})$
have two types of stationary solutions of finite amplitudes. The first type
corresponds to simple stripe solutions with only one finite modulus
\begin{align}
\label{rollsol}
|A_1|=\sqrt{\frac{\eta}{\gamma}}~,\, |A_2|=0 \quad {\rm or} \quad
|A_1|=0~,\, |A_2|=\sqrt{\frac{\eta}{\gamma}}~\,.
\end{align}

For the second type of solutions the amplitudes have identical moduli
\begin{eqnarray}
\label{squaresol}
|A_1|=|A_2|= \sqrt{\frac{\eta}{\gamma+\chi}}
\end{eqnarray}
which corresponds to a square pattern as can be seen from Eq.~(\ref{squareansatz}).

\begin {figure}[hbt]
\includegraphics[width=0.95\columnwidth]{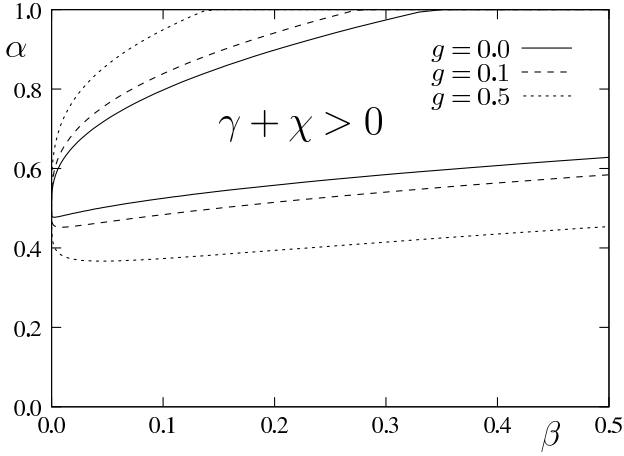}
\caption{
The nonlinear coefficient $\gamma+\chi$ as given by  Eq.~\eqref{nchi} is positive and square pattern 
bifurcate supercritically in the range enclosed by the solid line for $g=0.0$, for $g=0.1$ in the 
range enclosed by the dashed line and for $g=0.5$ between the dotted line. 
} 
\label{figbifsquare}
\end   {figure}

If the sum of the two nonlinear coefficients is positive, i.e.  $\gamma +\chi>0$, 
a  square pattern  bifurcates supercritically from the  homogeneous state. 
A vanishing sum $\gamma +\chi=0$ marks the tricritical line of the 
square pattern, the bifurcation changes from a supercritical to a subcritical one.
The tricritical line is displayed for different values
of the interaction parameter $g$ in Fig.~\ref{figbifsquare}, where increasing values of
$g$ broaden also the range of supercritically bifurcating squares
in the $\alpha-\beta$ plane. 

\subsubsection{Hexagonal patterns}
\label{hexpatt}
In two-dimensional systems close to the threshold and without an up-down symmetry $N,V \to -N,-V $
as in the Eqs.~(\ref{scaleq}) hexagonal structures are often preferred in some parameter range
to stripe or square patterns \cite{CrossHo}.
In this section the amplitude equations of hexagons are presented as they result by a 
perturbation calculation from Eqs.~(\ref{scaleq}) similar as described in App.~\ref{appA}
for stripes.

Close to threshold a  hexagonal pattern can be described by a superposition 
of three plane waves (stripe patterns) as given by Eq.~(\ref{fieldamp1}), 
but where the three wave vectors  enclose an angle of $ 2 \pi / 3$
with respect to each other. 
The solution may thus be represented by
\begin{align}
\label{hexansatz}
\vec{u}_0=\evec{e}_0 \left(A_1 e^{i \vec{k}_1 \vec{r}} + A_2 e^{i \vec{k}_2 \vec{r}} + A_3 e^{i \vec{k}_3 \vec{r}}\right)  + c.c. \, ,
\end{align}
whereas the wave vectors $\vec{k}_i$  ($i=1,2,3$) 
are given by
\vspace{-3mm}
\begin{align}
\vec{k}_1     = k_c           \left(\begin{array}{c}  1 \\ 0 \end{array} \right)
\quad{\rm and}\quad
\vec{k}_{2,3} = \frac{k_c}{2} \left(\begin{array}{c} -1 \\ \pm  \sqrt{3} \end{array}\right)\,.
\end{align}
The coupled amplitude equations for the three envelope functions $A_i$ ($i=1,2,3$) 
are of the following form
\begin{subequations}
\label{amplihex}
\begin{align}
\tau_0 \partial_t A_1 =& \,\eta\, A_1 + \delta\, A_2^* A_3^* \nonumber \\
 &\quad -\left( \gamma\, |A_1|^2 + \rho\, |A_2|^2 + \rho\, |A_3|^2 \right)A_1\,, \\
\tau_0 \partial_t A_2 =& \,\eta\, A_2 + \delta\, A_3^* A_1^* \nonumber \\
 &\quad -\left( \gamma\, |A_2|^2 + \rho\, |A_3|^2 + \rho\, |A_1|^2 \right)A_2\,, \\
\tau_0 \partial_t A_3 =& \,\eta\, A_3 + \delta\, A_1^* A_2^* \nonumber \\
 &\quad -\left( \gamma\, |A_3|^2 + \rho\, |A_1|^2 + \rho\, |A_2|^2 \right)A_3\,,
\end{align}
\end{subequations}
with $A_i^*$ being the complex conjugate of $A_i$.
$\tau_0$ and $\gamma$ are defined by the same expressions as for stripes and squares above 
in Eq.~\eqref{431} and Eq.~\eqref{ngamma} and the two nonlinear coupling constants $\delta$ and $\rho$ 
read within the scope of our model system
\begin{subequations}
\begin{align}
\label{deltadef}
\delta =& \frac{1+\sqrt{\beta}-2 \alpha}{1+ \sqrt{\beta}}\,,\\
\label{rhodef}
\rho =& \frac{6g}{1+\sqrt{\beta}} - \frac{4\alpha^2}{(1+\sqrt{\beta})^3} +
\frac{2\alpha(1+\alpha)}{(1+\sqrt{\beta})^2} -
\frac{3(2+\sqrt{\beta})}{4(1+\sqrt{\beta})}  \nonumber\\
&- \frac{(1-2\alpha)(3+2\alpha)}{4\sqrt{\beta}(1+\sqrt{\beta})}
+ \frac{2\alpha(1-2\alpha)}{\sqrt{\beta}} +3\alpha\,.
\end{align}
\end{subequations}
These three coupled nonlinear equations  (\ref{amplihex})
may again be derived via the relation
\begin{align}
\tau_0 \partial_t A_i = - \frac{\delta \pot{F}_H}{\delta A_i^* }
\end{align}
from a functional
\begin{align}
\label{pothexa}
\pot{F}_H =& 
 \sum_{i=1}^3 \left( \frac{\gamma}{2}\,|A_i|^4 - \eta\,|A_i|^2 \right)
+ \frac{\rho}{2} \sum_{i \neq j}^3 |A_i|^2\, |A_j|^2 
\nonumber \\
& \qquad\qquad
- \delta\,\left( A_1 A_2 A_3 + A_1^* A_2^* A_3^* \right)\,.
\end{align}
Eqs.~(\ref{amplihex}) admit two types of homogeneous solutions. The first one corresponds to a 
stripe solution with only one non-vanishing amplitude. For hexagonal solutions the moduli
of the three amplitudes $|A_1|=|A_2|=|A_3|=A$ coincide, but if one allows still a relative phase 
shift $\phi_i$ ($i=1,2,3$), with 
\begin{align}
A_i = A \, e^{i\phi_i}\,,
\end{align}
one obtains the nonlinear equation
\begin{eqnarray}
\label{noneqA}
0= \eta A + \delta\, e^{i\Phi} A^2 - (\gamma + 2 \rho) A^3\,,
\end{eqnarray}
with the  sum of the three phase angles  $\Phi = \phi_1+\phi_2+\phi_3$.
There are two real solutions of Eq.~\eqref{noneqA}
\begin{equation}
\label{hexamp}
A_{\pm} = \frac{1}{2(\gamma +2 \rho)} \left[ \delta \pm
\sqrt{ \delta^2  + 4 \eta \left(\gamma + 2 \rho \right)\,\,} \right]\,,
\end{equation}
with $A_+$ corresponding  to the larger amplitude for $\delta >0$ and $A_-$ for $\delta<0$.
For $\delta >0$ the phase angle is $\Phi=0$, which corresponds to 
{\it regular hexagons} and for $\delta <0$ the angle is $\Phi=\pi$,
which corresponds to  {\it inverse hexagons}.
Comparing for both solutions the functional $\pot{F}^\pm$ 
given by Eq.~\eqref{pothexa}, one 
finds that regular hexagons have the lower functional, i.e. 
$\pot{F}_H^+< \pot{F}_H^-$, in the range with $\delta>0$ 
and inverse hexagons in the range of $\delta<0$ with   $\pot{F}_H^+> \pot{F}_H^-$.

\begin {figure}[hbt]
\includegraphics[width=0.95\columnwidth]{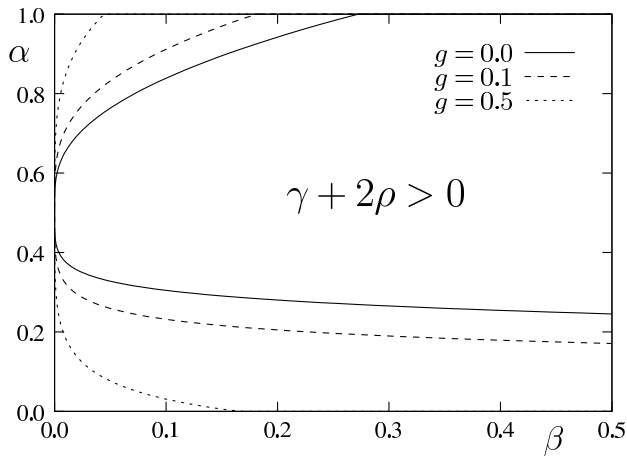}
\caption{The nonlinear coefficient  $\gamma+2\rho$ for
hexagons   is positive between the solid line for
$g=0.0$, for  $g=0.1$ between the dashed line and
 for $g=0.5$ between the dotted line. In each range the 
amplitude of the hexagonal solution  is limited by the
cubic terms in Eq.~\eqref{amplihex}.
}
\label{figbifhex}
\end   {figure}

The bifurcation from the homogeneous distribution of ion channels 
to a hexagonal modulation of the channel density
is subcritical according to the quadratic
nonlinearity $A^2$ in
Eq.~(\ref{noneqA}), which originates from
the quadratic nonlinearity $A^\ast_i A^\ast_j$ in Eqs.~(\ref{amplihex}).
However, the amplitudes $A_i$ are
still bounded by cubic nonlinearities in the parameter range of a 
positive   nonlinear coefficient 
$\gamma + 2 \rho>0$ in Eq.~(\ref{noneqA}). 
This nonlinear coefficient vanishes along the lines shown for 
different values of the parameter $g$ in the $\alpha-\beta$ plane 
in Fig.~\ref{figbifhex}.  
If this coefficient becomes negative, i.e. $\gamma + 2 \rho<0$, Eqs. (\ref{amplihex}) do not 
have any stationary, finite amplitude solutions.
In this case one needs either a higher order expansion or the amplitudes of hexagons
have to be determined by solving the basic equations (\ref{scaleq}) numerically.
Increasing values of the nonlinear interaction parameter $g$ enlarges the parameter range
wherein stationary solutions of the form \eqref{hexamp} occur.

\subsection{Competition between patterns}
\label{compet}

\begin {figure}[hbt]
\includegraphics [width=0.95\columnwidth] {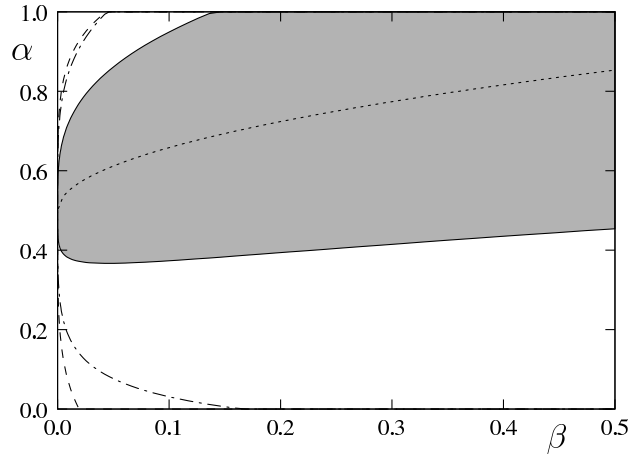}
\caption {
The bifurcation behavior of stripes, squares and hexagons is shown in the $\alpha$-$\beta$ 
plane for $g=0.5$:
In the shaded range the amplitudes for stripes, squares and hexagons are limited by a cubic
nonlinearity, i.e. $\gamma>0$, $\gamma+\chi>0$ and $\gamma+2 \rho>0$. On the right hand side of the
dashed line, which is determined by $\gamma=0$, stripes bifurcate supercritically and squares
do so in the range inclosed by the solid line, which is defined by $\gamma+\chi=0$. 
Between the dashed-dotted line which is determined by $\gamma+2\rho=0$ hexagons are limited
by cubic order terms.
}
\label{figbifall}
\end{figure}

\setcounter{paragraph}{0}
In the shaded subrange in Fig.~\ref{figbifall} stripes and squares bifurcate both supercritically 
and the amplitude of the hexagons is simultaneously limited by a cubic term. 
This range becomes even larger with increasing values of $g$ as indicated by the ranges in 
Fig.~\ref{figbif}, Fig.~\ref{figbifsquare} and Fig.~\ref{figbifhex}.
So the interesting question arises, which of the three solutions is preferred in this overlapping range.

One criterion is the comparison of the values of the functionals, i.e. to which solution belongs the lowest value 
of the respective functional $\pot{F}$.
The second criterion is the linear stability of each of the nonlinear solutions, i.e. in which subrange of 
the overlap range of parameters becomes one of the solutions linear unstable with respect to small perturbations.

\subsubsection{Comparison of the functionals for stripes, squares and hexagons.}
\begin {figure}[hbt]
\includegraphics[width=0.95\columnwidth]{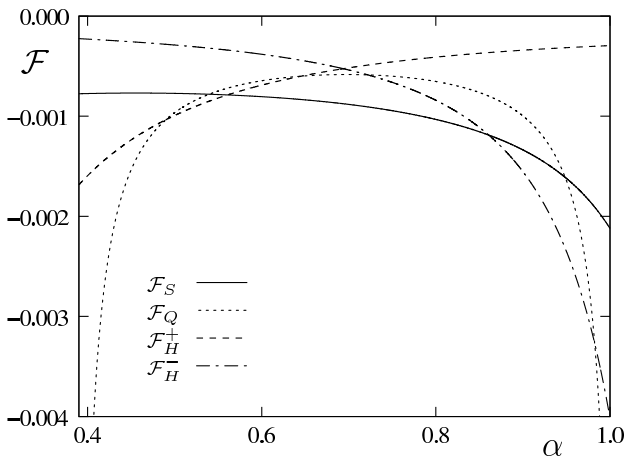}
\caption {
The functional $\pot{F}$ per unit area  is shown for stripes ($\pot{F}_S$),  
squares  ($\pot{F}_Q$),
regular ($\pot{F}_H^+$) and inverse hexagons  ($\pot{F}_H^-$)
as a function of the
parameter $\alpha$ as well as for the parameter values $\beta = 0.15$, 
$\eta = 0.06$ and $g=0.5$.
}
\label{figpot}
\end{figure}

For one set of parameters,  $\beta = 0.15$, $\eta = 0.06$ and $g=0.5$, the functionals 
of the three patterns are  shown in Fig.~\ref{figpot} as a function of $\alpha$ 
in the range where each of them bifurcates supercritically. This figure
indicates, in which region the respective pattern has the lowest value of the 
functional.

For small ($\alpha<0.49$) and very large values of $\alpha$ ($\alpha>0.98$) the functional
$\pot{F}_{Q}$ has the lowest value, i.e. squares have the lowest energy and are accordingly 
preferred. 
Regular hexagons $\pot{F}_H^+$ are preferred in the range $0.49<\alpha<0.56$, stripes 
$\pot{F}_{S}$ in the range $0.56<\alpha<0.86$ and finally inverse hexagons $\pot{F}_H^-$ 
in the range $0.86<\alpha<0.98$. 
These respective ranges change as a function of $\eta,\,\beta$ and $g$.

\begin{figure}[hbt]
\includegraphics[width=\columnwidth]{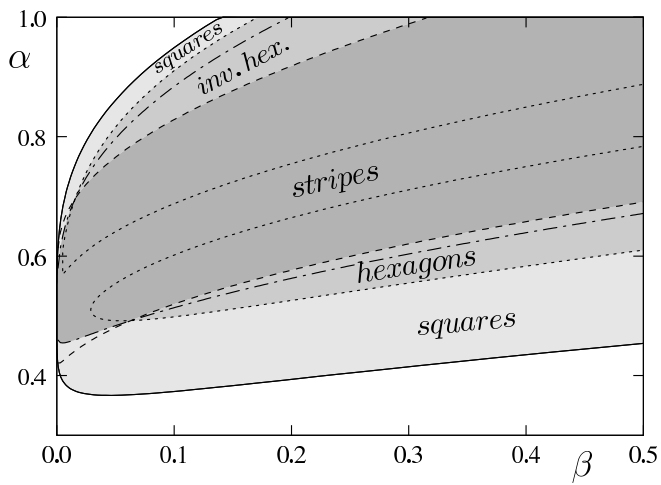}
\caption{
For $\eta=0.6$ and $g=0.5$ the parameter ranges are shown where
stripes (dark), hexagon (medium) and squares (bright) have the
lowest value for the functional $\pot{F}$.
Along the dotted line one has   $\pot{F}_{H}=\pot{F}_{Q}$, 
along the dashed line  $\pot{F}_{H}=\pot{F}_{S}$,
and along the dash-dotted line 
 $\pot{F}_{S}=\pot{F}_{Q}$.
}
\label{phasediag}
\end{figure}

Plotting the crossing points of the curves in Fig.~\ref{figpot} as a function of the kinetic 
parameter $\beta$ leads to a phase diagram as presented in Fig.~\ref{phasediag} for $\eta=0.06$ 
and $g=0.5$.
In this figure hexagons have a lower functional value than stripes beyond the upper dashed line
and below the lower dashed line and a lower functional value than squares between the dotted lines.
Taking the competition between squares and stripes into account too, stripes are preferred in the
dark shaded range, squares in the bright shaded range and hexagons in the medium shaded range.

Inserting the solutions of stripes in Eq.~\eqref{stripesol} and squares in Eq.~\eqref{squaresol} 
into their functionals in Eq.~\eqref{potstripe2} and Eq.~\eqref{potsquare} the functionals 
can be reduced to very simple expressions
\begin{align}
\pot{F}_S = -\frac{1}{2\gamma} \eta^2, \quad
\pot{F}_Q = -\frac{1}{\gamma+\chi} \eta^2\,.
\end{align}
Hence the comparison of the functionals for squares and stripes is independent of the reduced
control parameter $\eta$:
\begin{align}
\label{compareSQ}
\pot{F}_Q > \pot{F}_S 
\quad \Leftrightarrow \quad
\chi > \gamma>0\,.
\end{align}

Accordingly the curves, cf. the dash-dotted line in Fig.~\ref{phasediag}, as calculated
from the condition $\gamma=\chi$ of equal functional values, separate the regions where
the functionals of stripes or squares have the lower values.

\begin {figure}[hbt]
\includegraphics[width=\columnwidth]{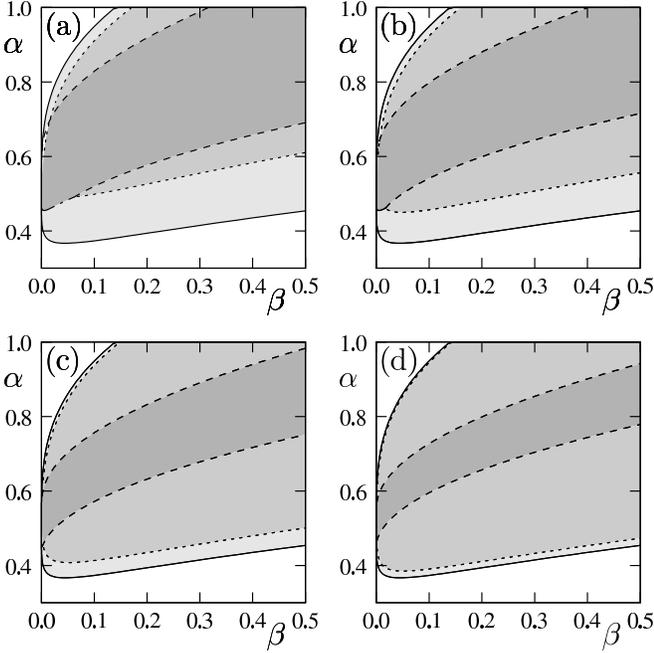}
\caption{
Dependence of the phase diagram of square, hexagon and stripe solutions
(bright, medium and dark gray) on the control parameter $\eta$:
(a)-(d) $\eta=0.6, 0.4, 0.2, 0.1$, $g=0.5$.
For small values of $\eta$ the region for hexagon solutions in the 
$\alpha$-$\beta$-plane increases and other solutions, squares quite
more than stripe patterns, are suppressed. 
}
\label{phasediag4er}
\end{figure}

However, a comparison with the functionals for regular and inverse hexagons is 
not independent 
of $\eta$.
Since hexagons bifurcate subcritically their amplitude is already finite and they have lower 
functional values at threshold $\eta=0$, i.e. $\pot{F}_H(\eta=0)<\pot{F}_S=\pot{F}_Q=0$. 
Hexagons are therefore always preferred close to the threshold.
Stripes and squares are always favored with respect to hexagons beyond some critical 
values $\eta>\eta_S(\alpha,\beta)$ and $\eta>\eta_Q(\alpha,\beta)$ which are determined by the 
conditions $\pot{F}_H=\pot{F}_S$ and $\pot{F}_H=\pot{F}_Q$, respectively.
Accordingly, with decreasing values of $\eta$ the range in the $\alpha-\beta$ plane increases
where hexagons have the lowest functional. As indicated by Fig.~\ref{phasediag4er} the ranges
of stripes and squares become smaller and smaller. For small values of $\eta$ square patterns
are suppressed nearly completely.

For the interesting limiting case, $\alpha =1/2$ and $\beta \to 0$, one has $\delta =0$ and $\rho = 6 g + 1/2$.
Thus one expects in two spatial dimensions for $\beta \ll 1$ and $\alpha= 1/2$ as well as small values of 
$\eta$ a clustering of ion channels to stripes.
For finite values of $\beta$ and $\eta >0$ stripes are preferred to hexagons in the neighborhood of the curve
$\delta=0$, cf. Fig.~\ref{figbifall}. This region broadens with increasing values of $\eta$, as indicated by 
Fig.~\ref{phasediag4er}.

\subsubsection{Linear stability analysis}
\setcounter{paragraph}{0}
A comparison of the values of the functionals for the respective patterns is one criterion for 
their basin of attraction. A linear stability analysis of the patterns shows, as described subsequently,
that two patterns can both coexist in a region around the parameter range where their 
functionals agree.

\paragraph{Stripes vs. hexagons.}
Stripes are still linearly stable even if their free energy is already higher than that of the hexagons, 
as can be seen by investigating the linear stability of stripes with respect to small amplitude
perturbations $\delta A_i$
\begin{align}
A_1=  \sqrt{\frac{\eta}{\gamma}}+ \delta\! A_1, \qquad A_2=\delta\! A_2, \qquad A_3 = \delta\! A_3 \,.
\end{align}
Linearizing Eqs.~\eqref{amplihex} with respect to the small functions $\delta\! A_i(t) $ and solving 
the resulting linear differential equations, one obtains the stability boundary as described by the condition (see e.g. Ref.~\cite{Ciliberto:90.1})
\begin{align}
\label{stablestripe}
\eta\left(\gamma-\rho\right)^2-\gamma\delta^2 >0\,.
\end{align}
By a similar stability analysis of the hexagonal solutions one obtains the following  stability boundary \cite{Ciliberto:90.1}
\begin{align}
\label{stablehexa}
\eta\left(\gamma-\rho\right)^2-\left(\gamma-2\rho\right)\delta^2<0\,.
\end{align}
While stripes have the lower free energy between the solid lines in
Fig.~\ref{phasediag}, stripes become only unstable
at the dashed line and the hexagons at the dash--dotted line in Fig.~\ref{phasediag}.

\paragraph{Stripes vs. squares.}
By a linear stability analysis of the stationary solutions given by Eq.~(\ref{rollsol}) and by
Eq.~(\ref{squaresol}) one finds that stable stripes are preferred in the range  $\chi> \gamma>0$ 
of the nonlinear coefficients and squares in the parameter range $|\chi| < \gamma$ \cite{Segel:65.1}.
These ranges coincide with the ranges where both patterns have their lower functional values
(cf. Eqs. \eqref{compareSQ}). Therefore stripe and square patterns do not coexist.

\paragraph{Squares vs. hexagons.}
Numerical results in Sec.~\ref{numSquare} show that the amplitude equation for squares is a good 
approximation only for very small values of $\eta$. 
In this range square patterns are nearly completely suppressed by hexagonal pattern. 
Thus the quite complicated stability analysis of square patterns vs. hexagons 
(see e.g. \cite{Kubstrup:96.1}) has been left out.

\begin {figure}[hbt]
\includegraphics[width=0.95\columnwidth]{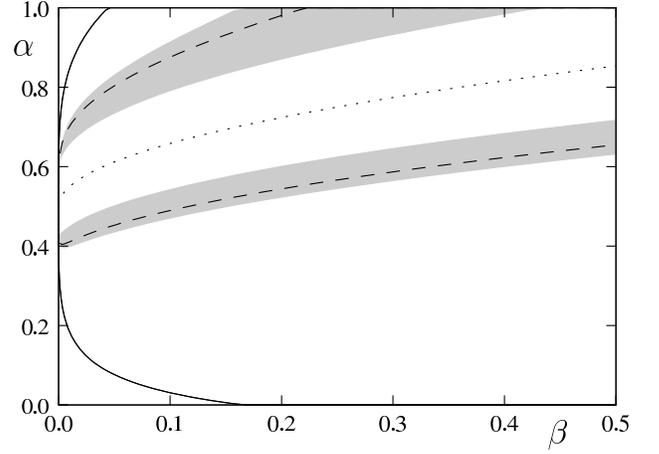}
\caption{
Range of coexistence of hexagons and stripes for $\eta=0.1$ and $g=0.5$. 
In the grey regions hexagons and stripes are both linearly stable with respect to small amplitude
perturbations. Therefore both patterns coexist in a finite range around the dashed lines where 
their functionals agree, $\pot{F}_S=\pot{F}_H$.
Along the dotted line the nonlinear coefficient $\delta(\alpha,\beta)=0$ vanishes.
The solid line indicates the region where both structures are limited by a cubic nonlinearity
in their amplitude equations ($\gamma>0$ and $\gamma+2\rho>0$).
}
\label{stability}
\end{figure}

\subsection{``Nearly conserved case'' -  $\beta \propto \eta^2$}
\label{weakspin}

The amplitude equations for stripes, squares and hexagons were
derived in Sec.~\ref{weakper} under the assumption of 
a finite wave number  $k_c =\beta^{1/4} \propto O(1)$ and
$k_c^2 \gg \eta$.  
In the limit of a conserved number of open ion channels, i.e.  $\beta=0$, beyond the onset
of patter formation a clustering of channels takes place \cite{Fromherz:1989.1},
that may be described with a single component Cahn-Hilliard equation.
In this section we show that in the limit of small values of the rate parameter $\beta$ with  $\beta \propto \eta^2$
a reduction of Eqs.~(\ref{scaleq}) to a single equation is still  possible, whereby the resulting equation
covers the qualitative properties of demixing as well as of periodic pattern formation. 

From the previous section it is known that in the limit $\beta \to 0$ the bifurcation is subcritical 
in the range of small values of $\beta$, besides $\alpha=1/2$. Therefore the appropriate expansion,
which leads also to finite amplitude solutions, is the combined expansion  of  $\beta \propto \eta^2$
and of  $\alpha=1/2+ \bar \alpha \sqrt{\eta}$ (with $\bar \alpha \propto O(1)$), as described 
in this section.

\setcounter{paragraph}{0}
\paragraph{Expansion in the regime $\beta \propto \eta^2$ and $\alpha$ arbitrary.}
In this  case we expand the two fields $N$ and $V$ with respect to powers of $\eta$ 
in the following manner

\begin{subequations}
\begin{align}
N =& \eta N_0 + \eta^2 N_1 + \ldots \,\,\, , \\
V =& \eta V_0 + \eta^2 V_1 + \ldots \,\,\,.
\end{align}
\end{subequations}
In addition we also introduce slow space variables, $X,Y = \sqrt{\eta } \,x, \sqrt{\eta}\,y $, and
a slow time variable $T = \eta^2 t$.
Choosing $\beta = \eta^2 \tilde \beta$, with $\tilde \beta \propto O(1)$ and expanding Eqs.~\eqref{scaleq}
with respect to powers of $\eta$ one obtains a single order parameter equation for $N_0$:
\begin{align}
\partial_T N_0 =& - \tilde \nabla^2\left[\left( \tilde \nabla^2  + 1 + 2\sqrt{\tilde \beta}\right)  N_0
- \left(\alpha - \frac{1}{2}\right)  N_0^2\right]            \nonumber \\
&- \tilde \beta N_0
\end{align}
wherein  $\tilde \nabla = (\partial_X, \partial_Y)$ has been used. 
The voltage follows immediately  via the identity
\begin{align}
V_0 =& - N_0 \,.
\end{align}
By rescaling space, time as well as the amplitude ($\tilde N = \eta N_0$) this equation takes the form
\begin{subequations}
\begin{align}
\label{norm1}
\partial_t \tilde N =& -\left( \nabla^2 +\eta +2\sqrt{\beta}\right) \nabla^2  \tilde N,
+ \left(\alpha - \frac{1}{2}\right) \nabla^2 (\tilde  N^2) \nonumber \\
& - \beta \tilde N \, , \\
\tilde  V =& - \tilde N\,.
\end{align}
\end{subequations}
This equation for $\tilde N$ shares similarities with the damped Kuramoto-Sivashinsky equation
\cite{Kuramoto:84,CrossHo}.
The only difference is in the nonlinearity
$ \nabla^2(\tilde N^2) = 2 ( (\nabla \tilde N)^2 + \tilde N \nabla^2 \tilde N)$
because the Kuramoto-Sivashinsky equation includes as nonlinearity  $(\nabla \tilde N)^2$ only.
The additional term, $\tilde N \nabla^2 \tilde N$, however, changes the dynamics and
stability  of the solutions completely beyond threshold, $\eta > 0$.
While one has for the Kuramoto-Sivashinsky equation ``turbulent'' but bounded solutions, 
the solutions of Eq.~(\ref{norm1}) are  always divergent according to the nonlinear diffusion
$\tilde N \nabla^2 \tilde N$.
For $\alpha = 1/2$ the nonlinear coefficient in  Eq.~\eqref{norm1} vanishes completely,
which suggests a  different expansion close to this point, as described in the next paragraph.

\paragraph{Expansion in the regime $\beta \propto \eta^2$ and $\left(\alpha-\frac{1}{2}\right) \propto \sqrt{\eta}$.}
Here also the parameter $\alpha$ is expanded with respect to
the small parameter $\eta$: $\alpha = 1/2 + \sqrt{\eta\,} \tilde \alpha$ with $\tilde \alpha \propto O(1)$.
Expanding the fields $N$ and $V$ with respect to powers of $\sqrt{\eta}$
\begin{subequations}
\begin{align}
N =& \, \eta^{\frac{1}{2}} N_0 + \eta N_1 + \eta^{\frac{3}{2}} N_2 + \dots \, ,\\
V =& \, \eta^{\frac{1}{2}} V_0 + \eta V_1 + \eta^{\frac{3}{2}} V_2 + \dots
\end{align}
\end{subequations}
yields from Eq.~\eqref{scaleq} at leading order a single equation for $N_0$, which
after rescaling $\tilde N = \sqrt{\eta\,} N_0$ takes the form
\begin{subequations}
\begin{align}
\label{Nb2}
\partial_t \tilde N =
& - \nabla^2 \left[\left( \nabla^2 +\eta + 2 \sqrt{\beta}\right) \tilde N
  - \left(\alpha-\frac{1}{2}\right)  \tilde N^2 \right. \nonumber \\
& \left.-\left( \frac{1}{12} + g\right) \tilde N^3  \right] - \beta \tilde N \,,\\
\tilde  V =
& - \tilde N\,.
\end{align}
\end{subequations}
The cubic nonlinearity now limits the amplitudes of the solutions to finite values, 
because $1/12 + g$ is positive even in the limit $\beta \to 0$.
For $\beta = 0$ and  $\alpha = 1/2 $, corresponding to $\tilde \alpha = 0$, the transition to the 
inhomogeneous channel distribution is continuous while it is discontinuous for $\tilde \alpha \neq 0$
but remains bounded.
In the limit of a conserved channel density $\tilde N$, i.e. $\beta=0$, equation \eqref{Nb2} 
is of the Cahn-Hilliard type \cite{Cahn:58.1,Lubensky:95}.

Eq.~\eqref{Nb2} covers all qualitative features of the 
basic equations (\ref{scaleq}) in both cases, in the limit $\beta=0$ and 
for $\beta \not =0$. Therefore, 
similar as in the previous section one can also  derive the amplitude equations 
for stripes, squares  and
hexagons by starting from the  modified equation (\ref{Nb2}) instead of Eqs.~\eqref{scaleq}. These
derivations are much simpler for Eq.~(\ref{Nb2})  but the results are now 
restricted to a range along the line $\alpha \sim 1/2$,  close to the threshold
and to small values of the rate parameter $\beta \propto \eta^2$.
Thereby one obtains again the amplitude equation for stripes as given by 
Eq.~\eqref{amplitudeeq}, for squares as by Eq.~(\ref{amplisquare}) 
and for hexagons as by Eq.~\eqref{amplihex}, but now with slightly 
modified  expressions for the coefficients.
\begin{align}
\tau_0 &= \frac{1}{\sqrt{\beta}}, &  \delta &= 1-2\alpha,     \nonumber \\
\gamma &= \frac{1}{4}+3g,         &  \rho &= \chi =  \frac{1}{2}+6g. 
\end{align}
These expressions may also be recovered from the formulas given in Sec.~\ref{weakper}
in the limit $\beta \to 0$ and $\alpha \to 1/2$.

\section{Numerical Results}
\label{numresults}

The amplitude equations, as given for the present system in the previous sections, are exemplarily derived 
in the appendix by a perturbational calculation. However, the validity range of these equations and their 
solutions is a priori unknown.
An estimation of this range can be provided by comparing the analytical solutions with 
numerical simulations of the basic equations \eqref{scaleq}, as done in this section.

\subsection{Stripe patterns}
\begin {figure}[hbt]
\includegraphics[width=0.95\columnwidth]{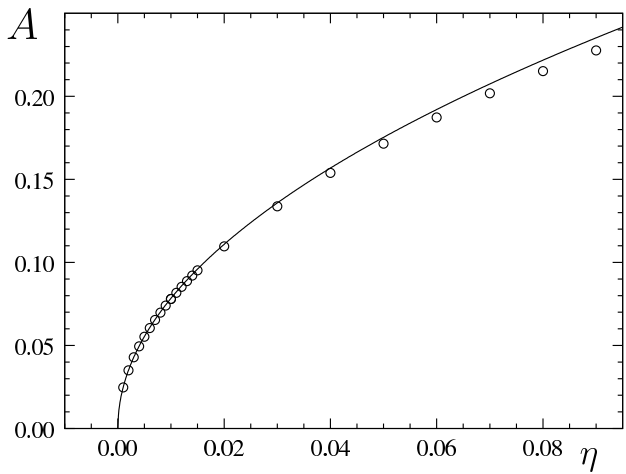}
\caption {
The amplitude 
of a  stripe pattern as determined by the
amplitude equation (solid line) and by numerical solutions of Eq.~\eqref{scaleq} (circles) 
are compared as a function of the control parameter. For the parameters 
$\alpha=0.4,~\beta=0.1,~g=0.5$ and $\tau_V=0.1$ the bifurcation
was supercritical, i.e.  $\gamma>0$.
}
\label   {welleich}
\end   {figure}

In the range of supercritically bifurcating stripe patterns determined analytically in the
previous section, the analytical and numerical solutions of Eqs.~\eqref{scaleq} are compared in
Fig.~\ref{welleich} as a function of the control parameter $\eta$ for  $\alpha=0.4,
\,\beta=0.1,\,g=0.5$ and $\tau_V=0.1$.
There is a fairly good agreement between the analytical and the numerical solutions up to about
$\eta=0.1$.
Since the stripe pattern is  stationary it does not depend on the actual value of $\tau_V$ used
in simulations.

\subsection{Square patterns}
\label{numSquare}
\begin {figure}[hbt]
\includegraphics[width=8cm]{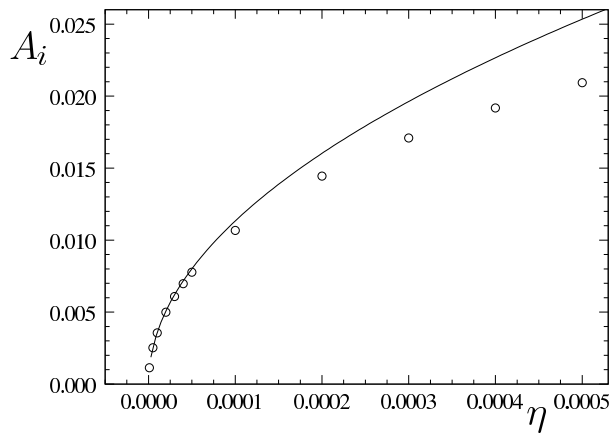}
\caption {
The amplitudes of squares, $A_1=A_2$, are shown as a function of the reduced control parameter $\eta$
and for the parameters $\alpha=0.4$, $\beta=0.1$, $g=0.5$.
The solid line is given by Eq. \eqref{squaresol} and the circles are obtained from numerical simulations of the
model equations.
}
\label{squareAmp}
\end{figure}

Numerical simulations do not show stable square patterns besides transients to 
finally hexagonal patterns.
Choosing a special geometry with a very small system length ($L=2\pi/k_c$) hexagonal patterns
can be suppressed in numerical simulations. In this case the system shows the square patterns 
as predicted. 
By comparing the amplitudes $A_i$ as given for squares by Eq.~\eqref{squaresol} for $\alpha=0.4$, 
$\beta=0.1$ and $g=0.5$ with the numerical obtained solution, as depicted in Fig.~\ref{squareAmp}, we find 
as a function of the reduced control parameter only an acceptable agreement for very small values of 
$\eta \lesssim 10^{-4}$.
But then we find that squares are only preferred according to the analytical calculation
in the range $\eta \gtrsim 0.1$, which is far beyond the validity range of the amplitude equations
for squares. This is an explanation why we do not find the predicted squares by numerical solution
of Eqs.~\eqref{scaleq}.

\subsection{Hexagonal patterns}
\begin {figure}[hbt]
\includegraphics[width=0.95\columnwidth]{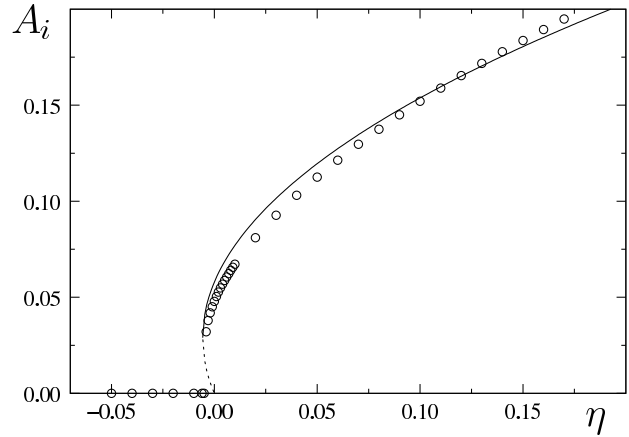}
\caption 
{The amplitudes $A_j$ of the hexagonal pattern
are given as a function of the reduced control parameter $\eta$. 
The solid line corresponds to the analytical solution
$A_{+}$ 
and the dashed line 
belongs to the unstable solution $A_-$, 
where both are given  
by  Eq.~\eqref{hexamp}.
The data points are obtained from the numerical
solution of Eqs.~\eqref{scaleq}.
The parameters
$\beta=0.1$, $\alpha=0.4$ and $g=0.5$ have been used.
}
\label   {fighexamp}
\end   {figure}

Close to threshold hexagons are the
preferred pattern in a wide range of parameters.
In a range where stripes and squares bifurcate supercritically,
but where hexagons are already preferred, at $\alpha=0.4$,
$\beta=0.1$ and $g=0.5$, the analytically, cf. Eq.~\eqref{hexamp},
 and the numerically obtained solutions are compared in
Fig.~\ref{fighexamp}.

\subsection{Anharmonic solutions and clustering of ion channels for near conservation}
\begin {figure}[hbt]
\includegraphics[width=\columnwidth]{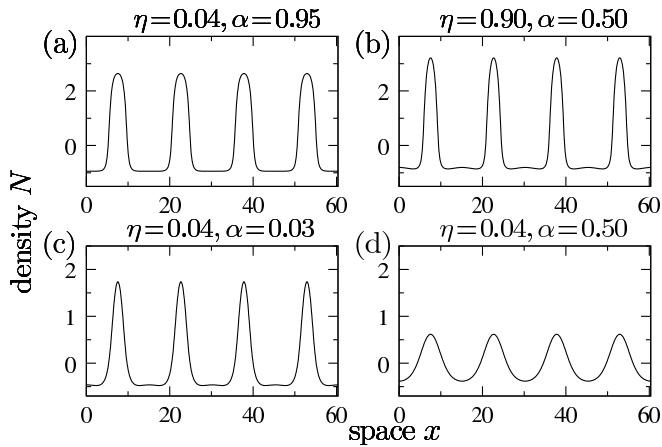}
\caption {The stationary and normalized distribution of ion channels 
$N(x)$ is shown for a stripe pattern above threshold 
for  four different
sets of parameters. In all cases 
$\beta=0.03$ and $g=0.1$ were fixed and therefore also the critical wave number
 $k_c=0.416$. The system length is $L = 8\pi/k_c$.
}
\label{oneplate}
\end{figure}

Increasing the control parameter up to $\eta=0.9$, far beyond the validity range
of the amplitude equations, the density $N(x)$ becomes rather anharmonic as
shown in Fig.~\ref{oneplate}(b). 
Closer to the threshold at
 $\eta=0.04$ but in the range where stripes bifurcate
subcritically and where the amplitudes take immediately
large values, at $\alpha=0.95$ and $\alpha=0.03$,
the solutions are also very  anharmonic as shown for
 $N(x)$ in Fig.~\ref{oneplate}(a) and \ref{oneplate}(c).
In both cases each peak in Fig.~\ref{oneplate}(a) and (c)
takes already a similar shape that are typical for
the clusters in the conserved limit $\beta =0$.

\section{Discussion and Conclusion}
\label{conclusion}

In Ref.~\cite{Zimmermann:95.2} a model for the dynamics of ion channels
including electrophoresis, an opening-closing reaction as well as a
simultaneous binding-release reaction has been introduced.
Compared to this earlier work, we 
have taken into account the effect of the
excluded volume interaction between ion channel molecules.
In addition the analysis of the bifurcations beyond 
the threshold of pattern formation
has been extended to two spatial dimensions. 

In terms of amplitude equations we give 
a detailed analysis of the competition between stripes, 
squares and hexagonal
patterns. We have found that
immediately above threshold hexagonal patterns are preferred
in a large range of parameters, whereas further beyond the threshold 
stripes are preferred 
in an increasingly larger parameter range. 

The validity range of the amplitude expansion has been
tested by solving 
the model equations numerically with a
pseudo-spectral code. By this
comparison we have shown that the 
amplitude equations for squares have a very small validity range.
The expansion for squares breaks down
two orders of magnitude earlier than
the one for stripes and hexagonal structures. 
Accordingly, the large range where square patterns should
be favored as 
predicted by the amplitude expansion is not confirmed
by the numerical analysis of the model equations
while the amplitude equations describing the
competition between stripes and hexagons
are in a fairly good agreement with the
numerical simulations close to the threshold.

In the limiting case $\beta \rightarrow 0$ implying a conserved number of open 
ion channels the patterns have a strong similarity with those patterns 
occurring during ion channel clustering \cite{Fromherz:1989.1} in systems
with a binding-release reaction.
Near threshold and in the limit of $\beta \propto \eta^2$
the model equations can be reduced to a single
model equation which shares similarities
with different variants of the Swift-Hohenberg equation
\cite{CrossHo} as well as the Cahn-Hilliard equation 
\cite{Cahn:58.1,Lubensky:95}. On the basis of this reduced equation
essential effects related to the binding release reaction are
already captured.

It is an interesting question whether the curvature of membranes 
influences the pattern competition especially between
stripes and hexagons. 
One expects that in such a system the effects of a broken up-down symmetry 
become stronger leading to an additional enlargement of the parameter 
range where hexagonal patterns occur. 

If the binding-release reaction is replaced by
an opening-closing reaction, the formation of spatially
periodic either stationary or oscillatory patterns 
has been reported \cite{SKramer:2002.1,RPeter:2006.100}. 
There the nonlinear behavior of stripe patterns 
has been discussed 
only partially and only in one spatial dimension.
A competition of patterns as described in the present work will be expected 
but also a competition between two-dimensional stationary and time-dependent 
patterns.

Our calculations are related to experiments  of the
type as investigated in Ref.~\cite{Rentschler:1998a}
where ion channels are studied in vitro.
The electro-diffusive model at hand seems to be relevant for in vivo
systems as shown by experiments on the effect of electric fields on
clustering of acetylcholine receptors \cite{Leonetti:2005.1,Stollberg:88.1, Nitkin:90.1}.
In membranes composed of several types of lipids
a selforganized structuring has also been described
in terms of lipid rafts \cite{Simons:1997a}. Therefore one expects in such cases
a spatially varying mobility of proteins embedded in the membrane. How
this heterogeneity affects the formation of patterns
is another interesting question as has been investigated for other model systems
\cite{Zimmermann:1993a,Schmitz:1996a,Epstein:2001.2,Peter:2005a,Hammele:2006.1}.
A detailed analysis of all these questions may be given in forthcoming works.

\begin{acknowledgments}
We would like to thank Falko Ziebert, Ronny Peter and Ernesto Nicola for fruitful discussions.
\end{acknowledgments}

\appendix

\section{Derivation of the amplitude equation for the stripe pattern.}
\label{appA}
\subsection{Basic equations  in Matrix-Notation}
Rewriting the basic Eqs.~(\ref{scaleq}) to a matrix notation
\begin{equation}
\label{233}
(\mat{M} \cdot \partial_t + \mat{L}) \vec{u} = \vec{N}(\vec{u})
\,,
\end{equation}
allows a more compact formulation of the derivation of
the generic amplitude equations
of stripe patterns as given by  Eq.~\eqref{amplitudeeq},
of square patterns as given by Eqs.~\eqref{amplisquare} or 
of hexagonal patterns as given by Eqs.~\eqref{amplihex}.
The two components of the vector $\vec{u}$ are the  
normalized channel density  $N$ and  
the reduced voltage $V$, whereas the matrices $\mat{M}$ and $\mat{L}$ 
represent the linear parts of Eqs.~(\ref{scaleq}) 
\begin{align}
\label{235}
\mat{M} &= \left( \begin{array}{ccc} 1 & 0  \\ 0 & 0 \\
\end{array} \right) \, ,&
\mat{L} &= \left( \begin{array}{ccc}
\beta - \nabla^2 \,\,\,\,,  & - \nabla^2\\
\alpha ( 1- \alpha) \varepsilon\,,  &  1 - \nabla^2
\end{array} \right)\, ,
\end{align}
and the vector $\vec{N}$ the nonlinearity:
\begin{align}
\label{234}
\vec{N} &=
\left( \begin{array}{c} 
\nabla \left( N \nabla V \right)
+ g \nabla^2 (N^3) \\ -\alpha N V
\end{array} \right) \, . 
\end{align}
The linear coefficients of Eq.~(\ref{amplitudeeq}) follow directly from the
linear stability analysis as described in Sec.~\ref{onset} and Sec.~\ref{onedipat}, but the
nonlinear coefficient $\gamma$ in the same equation is determined by the
perturbation expansion described in the next subsection.

\subsection{Nonlinear coefficient}
The sign of the  nonlinear coefficient $\gamma$ in Eq.~\eqref{amplitudeeq} 
determines, whether  one has a sub- or a supercritical  bifurcation to the 
periodic state given by  Eq.~(\ref{fieldamp1}). The scheme of the derivation 
of $\gamma$ and the respective amplitude equations may be found in various 
references, as for instance in Refs.~\cite{CrossHo,Newell:92.1,Zimmermann:93.1}.
This scheme  for the derivation of $\gamma$  is summarized for the present 
system  in this appendix.  The starting point is an expansion of the solutions 
of Eqs.~(\ref{233}) with respect to powers of the reduced control parameter  
$\eta$ as given by Eq.~(\ref{eta}):
\begin{align}
\label{404}
\vec{u}(\vec{r},t) =
\eta^{\frac{1}{2}} \vec{u}_0(\vec{r},t) +
\eta \,            \vec{u}_1(\vec{r},t) +
\eta^{\frac{3}{2}} \vec{u}_2(\vec{r},t) + \dots\,.
\end{align}
Since the vector  $\vec{N}$ is a nonlinear function of $\vec{u}$, it may be 
also expanded with respect to powers of $\eta$:
\begin{eqnarray}
\label{NonlinexpN}
\vec{N}(\vec{u}) = \eta\, \vec{N}_1(\vec{u}_0)  + \eta^{\frac{3}{2}}\, \vec{N}_2(\vec{u}_0,\vec{u}_1)  + \ldots \enspace.
\end{eqnarray}
To the ansatz in Eq.~(\ref{fieldamp1}) for a homogeneous stripe pattern 
\begin{align}
\label{Afieldamp1}
\vec{u}_0 = \evec{e}_0\,A \,e^{i \vec{k}_c \vec{r}}  + c.c.
\end{align}
a multiscale analysis in time is added 
\begin{align}
\partial_t \rightarrow \partial_t + \eta\, \partial_T
\end{align}
to account for the variation of the amplitude $A=A(T)$ on a very slow time scale 
$T=\eta \,t$. 
Together with the  relation  $\varepsilon=\varepsilon_c (1+\eta)$ finally
the basic equations in (\ref{233}) may be rearranged into a  powers series 
with respect to $\eta^{\frac{1}{2}}$ leading to the following hierarchy of 
equations:
\begin{align}
\label{420a}
&\eta^{\frac{1}{2}} &:& 
&\left(\mat{M} \partial_t + \mat{L}_0\right) \vec{u}_0 &= 0 \,,\\
\label{420b}
&\eta &:& 
&\left(\mat{M} \partial_t + \mat{L}_0\right) \vec{u}_1 &=
\vec{N}_1(\vec{u}_0) \,,\\
\label{420c}
&\eta^{\frac{3}{2}} &:& 
&\left(\mat{M} \partial_t + \mat{L}_0\right) \vec{u}_2 &=
- \left(\mat{L}_2 + \mat{M} \partial_T \right) \vec{u}_0+\vec{N}_2(\vec{u}_0,\vec{u}_1)
\end{align}
with the linear operators 
\begin{align}
\label{defL0}
\mat{L}_0 &= \left( \begin{array}{ccc}
\beta - \nabla^2 \,\,\,\,,  & - \nabla^2\\
\alpha ( 1- \alpha) \varepsilon_c\,,  &  1 - \nabla^2
\end{array} \right) \, , \\
\label{defL2}
\mat{L}_2 &= \left( \begin{array}{ccc} 0 & 0  \\
\alpha (1-\alpha) \varepsilon_c & 0 
\end{array} \right)
\end{align}
and the nonlinearities
\begin{align}
\vec{N}_1 &= \left(\begin{array}{c}
\nabla \left(N_0 \nabla V_0  \right) \\
- \alpha N_0 V_0
\end{array} \right)\, , 
\\
\label{defN2} 
\vec{N}_2 &= 
\left( \begin{array}{c}
 \nabla ( N_0 \nabla V_1 + N_1 \nabla V_0 )+g \nabla^2 N_0^3\\
-\alpha (N_0 V_1 + N_1 V_0) 
\end{array} \right).
\end{align}
Using the ansatz of Eq.~\eqref{Afieldamp1} the leading contribution
$\vec{N}_1(\vec{u}_0)$ of  $\vec{N}(\vec{u})$ has the explicit form
\begin{align}
\vec{N}_1
 & =- E_0
\left( \begin{array}{c} 
2 k_c^2 \, A^2 e^{2i \vec{k}_c \vec{r}}  \\
\alpha \left( A^2 e^{2i \vec{k}_c \vec{r}} +|A|^2 \right)
\end{array} \right) + c.c. \, .
\end{align}

The solution $\vec{u}_1$ of Eq.~(\ref{420b}) 
has to be of the same form as the inhomogeneity $\vec{N}_1$. 
This leads to the ansatz
\begin{align}
\label{424}
\vec{u}_1 =& \left(B_1 \evec{e}_0+B_2 \evec{e}_1 \right)  A^2 e^{2 i \vec{k}_c \vec{r}} 
+\left(B_3 \evec{e}_0+B_4 \evec{e}_1 \right) |A|^2 +c.c. 
\end{align}
using the two eigenvectors $\evec{e}_{0,1} = \left(1, E_{0,1} \right)$ of $\mat{L}_0$ 
with
\begin{align}
E_0 &= -\left(1+\sqrt{\beta}\right)\,,
&E_1 &= \frac{1+\sqrt{\beta}}{\sqrt{\beta}}\, .
\end{align}
After inserting of \eqref{424} in \eqref{420b} one obtains by comparison of coefficients:
\begin{align}
\label{425} 
B_1 &= \frac{1}{9}\left(10-\alpha+2\sqrt{\beta}+
        \frac{3\alpha}{\sqrt{\beta}(1+\sqrt{\beta})}
        +\frac{2-7\alpha}{\sqrt{\beta}}\right)\,, \nonumber\\
B_2 &= \frac{1}{3}\left(2\sqrt{\beta}
-\frac{\alpha\sqrt{\beta}}{1+\sqrt{\beta}}\right)\,, \nonumber\\
B_3 &= - \frac{\alpha\sqrt{\beta}}{1+\sqrt{\beta}} \,,  \hspace{1cm}
B_4 = -B_3 \,. 
\end{align}
At the  next higher order, Eq.~(\ref{420c}), one has to deal with the second order correction 
of the linear operator, $\mat{L}_2 $, and of the vector $\vec{N}_2$ 
as defined in Eq. (\ref{defL2}) and Eq. (\ref{defN2}).
It is not necessary to solve Eq.~(\ref{420c}) explicitly.
One can use Fredholm's alternative or one simply can take 
advantage of  the following property 
\begin{eqnarray}
\label{426a}
\langle \evec{f}_0 e^{i  \vec{k}_c \vec{r}}, \mat{L}_0  \vec{u}_2 \rangle 
=  \frac{1}{S} \int_S d\vec{r}  ~ e^{-i \vec{k}_c \vec{r}}\evec{f}_0^{\dagger}\mat{L}_0 \vec{u}_2(\vec{r},t)=0
\, \nonumber \\
\end{eqnarray}
of the left eigenvector
\begin{align}
\label{LEvect}
\evec{f}_0 = \left(  \begin{array}{c}
1 \\ F_0^\star
\end{array}
\right), \quad F_0 = -\frac{\sqrt{\beta}}{1+\sqrt{\beta}},
\end{align}
which spans the adjoint kernel of $\mat{L}_0$.
Since $\vec{u}_0$ and $\vec{u}_1$ have an explicit dependency only on the
time scale $T$ but not on $t$ the corresponding derivatives can be neglected. 
Accordingly  all the terms at the right hand side of Eq.~(\ref{420c})
  projected onto 
$\evec{f}_0 e^{i \vec{k}_c \vec{r}}$ also have to vanish:
\begin{align}
\label{426}
\langle\evec{f}_0 e^{i \vec{k}_c \vec{r}}, 
\vec{N}_2-\left(\mat{L}_2+\mat{M}\partial_T\right)\vec{u}_0 \rangle=0 
\, .
\end{align}
This provides  the solubility condition for the determination of 
the amplitude $A$. For this purpose the contributions to 
the expressions $ \mat{L}_2  \vec{u}_0 $ 
and $\vec{N}_2 $ which are proportional to
$e^{i \vec{k}_c \vec{r}}$ are collected.
According to the Fredholm's alternative we obtain after projection
\begin{align}
\gamma |A|^2 A - A + \tau \partial_T A = 0
\end{align}
with the nonlinear coefficient 
\begin{eqnarray}
\label{ngammaap}
\gamma &= &\frac{3 g}{1+\sqrt{\beta}} - \frac{1}{3} 
\Big[\, \frac{ 6 \alpha^2 -\bigl( \alpha -2(1+ \sqrt{\beta}) \bigr)^2}{1+\sqrt{\beta}}
 \nonumber \\ 
 && \hspace{-6mm}+ \frac{2}{3 \sqrt{\beta}} \bigl( 4 \sqrt{\beta} - 2 \alpha +1  \bigr) 
 \bigl( \sqrt{ \beta} - 2 \alpha +1   \bigr)
 \, \Big]\,,
\end{eqnarray} 
and the relaxation time
\begin{align}
\tau = \frac{1}{\sqrt{\beta}\left(1+\sqrt{\beta} \right)}\,.
\end{align}
The nonlinear coefficient depends on the rate parameter $\beta$, 
the density parameter $\alpha$ and on the parameter $g$ for the excluded volume interaction. 
The relaxation time of the pattern 
as well as the nonlinear coefficient  $\gamma$ both 
diverge for a frozen binding-release reaction ($\beta \rightarrow 0$). 
In this limit the wave number $k_c$ tends to zero and the assumptions made
for the derivation of the amplitude equation 
are not longer fulfilled.

\bibliographystyle{prsty}

\end{document}